\documentclass{article}
\title{Concept and Definition of Complexity}
\author{Russell K. Standish\\Mathematics and Statistics, UNSW} 
\usepackage{epsf,pstricks,pst-node,a4wide,apalike}

\def\citeyear(#1)#2{(#1)\nocite{#2}}

\begin{document}
\maketitle

\section*{Keywords}
Complexity, Emergence, Complex Systems, Information Theory, Graph
Theory, Occam's Razor


\begin{abstract}
The term {\em complexity} is used informally both as a quality and as
a quantity. As a quality, complexity has something to do with our
ability to understand a system or object --- we understand simple
systems, but not complex ones. On another level, {\em complexity} is
used as a quantity, when we talk about something being more
complicated than another.

In this chapter, we explore the formalisation of both meanings of
complexity, which happened during the latter half of the twentieth
century.
\end{abstract}

\section{Introduction: Is complexity a quality or a quantity?}

The term {\em Complexity} has two distinct usages, which may be
categorised simply as either a quality or a quantity. We often speak
of complex systems as being  a particular class of systems that are
difficult to study using traditional analytic techniques. We have in
mind that biological organisms and ecosystems are {\em complex}, yet
systems like a pendulum, or a lever are simple. Complexity as a {\em
quality} is therefore what makes the systems complex.

However, we may also speak of complexity as a quantity --- with
statements like a human being being more complex than a nematode worm,
for example. Under such usage, complex and simple systems form a
continuum, characterised by the chosen complexity measure.

Bruce Edmonds \citeyear(1999){Edmonds99} performed a comprehensive
survey of complexity measures as part of his PhD thesis, however it
has not been updated to include measures proposed since that time.
However, it remains the most comprehensive resource of complexity
measures available to date.

The aim of this chapter is not to provide a catalogue of complexity
measures, but rather to select key measures and show how they
interrelate with each other within an overarching information
theoretic framework.

\section{Complexity as a quantity}

We have an intuitive notion of complexity as a quantity; we often
speak of something being more or less complex than something
else. However, capturing what we mean by complexity in a formal way
has proved far more difficult, than other more familiar quantities we
use, such as length, area and mass. 

In these more conventional cases, the quantities in question prove to
be decomposable in a linear way, ie a 5cm length can be broken into 5
equal parts 1 cm long; and they can also be directly compared --- a
mass can be compared with a standard mass by comparing the weights of
the two objects on a balance.

However, complexity is not like that. Cutting an object in half does
not leave you with two objects having half the complexity overall. Nor
can you easily compare the complexity of two objects, say an apple and
an orange, in the same way you can compare their masses.  

The fact that complexity includes a component due to the interactions
between subsystems rapidly leads to a combinatorial explosion in the
computational difficulty of using complexity measures that take this
into account. Therefore, the earliest attempts at deriving a measure
simply added up the complexities of the subsystems, ignoring the
component due to interactions between the subsystems.

The simplest such measure is the {\em number of parts} definition. A
car is more complex than a bicycle, because it contains more
parts. However, a pile of sand contains an enormous number of parts
(each grain of sand), yet it is not so complex since each grain of
sand is conceptually the same, and the order of the grains in the pile
is not important. Another definition used is the {\em number of
distinct parts}, which partially circumvents this problem. The problem
with this idea is that a shopping list and a Shakespearian play will
end up having the same complexity, since it is constructed from the
same set of parts (the 26 letters of the alphabet --- assuming the
shopping list includes items like zucchini, wax and quince, of
course). An even bigger problem is to define precisely what one means
by ``part''. This is an example of the {\em context dependence} of
complexity, which we'll explore further later.

Bonner and McShea have used these (organism size, number of cell
types) and other {\em proxy} complexity measures to analyse complexity
trends in evolution \cite{Bonner88,McShea96}. They argue that all these
measures trend in the same way when figures are available for the same
organism, hence are indicative of an underlying organism complexity
value. This approach is of most value when analysing trends within a
single phylogenetic line, such as the diversification of trilobytes.

\section{Graph Theoretic Measures of Complexity}

Since the pile of sand case indicates complexity is not simply the
number of components making up a system, the relationships between
components clearly contribute to the overall complexity. One can start
by caricaturing the system as a {\em graph} --- replacing the
components by abstract {\em vertices} or {\em nodes} and relationships
between nodes by abstract {\em edges} or {\em arcs}. 

Graph theory \cite{Diestel05} was founded by Euler in the 18th century
to solve the famous K\"onigsberg bridge problem. However, until the
1950s, only simple graphs that could be analysed in toto were
considered. Erd\"os and Renyi \citeyear(1959){Erdos-Renyi59}
introduced the concept of a {\em random graph}, which allowed one to
  treat large complex graphs statistically. Graphs of various sorts
  were readily recognised in nature, from food webs, personal or
  business contacts, sexual relations and the Internet amongst others.
  However, it soon became apparent that natural networks often had
  different statistical properties than general random graphs. Watts
  and Strogatz \citeyear(1998){Watts-Strogatz98} introduced the {\em
    small world} model, which has sparked a flurry of activity in
  recent years to measure networks such as the Internet, networks of
  collaborations between scientific authors and food webs in
  ecosystems \cite{Albert-Barabasi01}.

Graph theory provides a number of measures that can stand in for
complexity. The simplest of these is the {\em connectivity} of a
graph, namely the number of edges connecting vertices of the graph. A
fully connected graph, however, is no more complex than one that is
completely unconnected. As connectivity increases from zero, a {\em
percolation threshold} is reached where the graph changes from being
mostly discontinuous to mostly continuous. The most complex systems
tend to lie close to the percolation threshold. Another graph measure
used is {\em cyclomatic number} of a graph, basically the number of
independent loops it contains. The justification for using cyclomatic
number as a measure of complexity is that feedback loops introduce
nonlinearities in the system's behaviour, that produce complex
behaviour.


\def\graphnodes#1{
\begin{pspicture}(3,2)
\cnodeput(1,0){A}{A}
\cnodeput(0,1){B}{B}
\cnodeput(1,2){C}{C}
\cnodeput(3,1.5){D}{D}
\cnodeput(2,.5){E}{E}
#1
\end{pspicture}
}

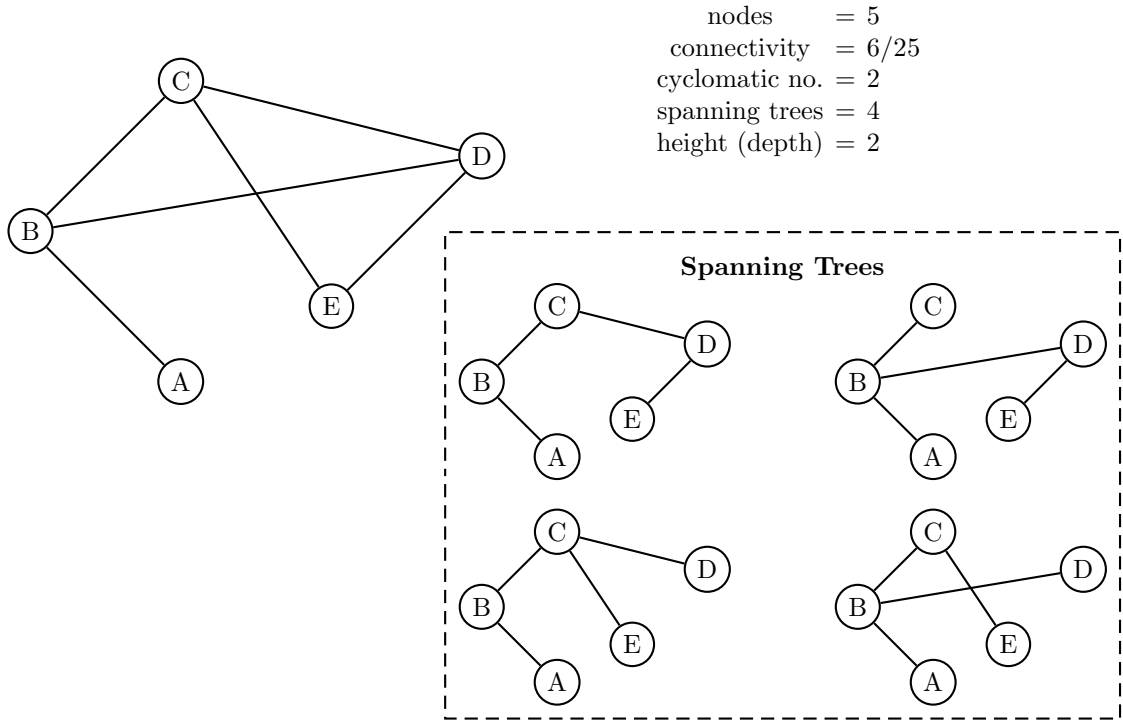
\begin{figure}
\begin{pspicture}(-.5,-.5)(14.5,9)

\rput[l](0,6)
{\psset{unit=2}
\graphnodes{
        \ncline{A}{B}
        \ncline{B}{C}
        \ncline{C}{D}
        \ncline{D}{E}
        \ncline{B}{D}
        \ncline{C}{E}
        }
}

\rput[l](8,8){
\begin{tabular}{c@{\hspace{1ex}=\hspace{1ex}}l}
nodes & 5\\
connectivity & 6/25\\
cyclomatic no. & 2\\
spanning trees & 4\\
height (depth) & 2\\
\end{tabular}
}

\psframe[linestyle=dashed](5.5,-.5)(14.5,6)
\rput(10,5.5){{\bf Spanning Trees}}
\rput(7.5,4){
  \graphnodes{
    \ncline{A}{B}\ncline{B}{C}\ncline{C}{D}\ncline{D}{E}
    }
  }
\rput(12.5,4){
  \graphnodes{
    \ncline{A}{B}\ncline{B}{C}\ncline{B}{D}\ncline{D}{E}
    }
  }
\rput(7.5,1){
  \graphnodes{
    \ncline{A}{B}\ncline{B}{C}\ncline{C}{D}\ncline{C}{E}
    }
  }
\rput(12.5,1){
  \graphnodes{
    \ncline{A}{B}\ncline{B}{C}\ncline{B}{D}\ncline{C}{E}
    }
  }

\end{pspicture}
\caption{Various graph theoretic measures for a simple graph. The
spanning trees are shown in the dashed box}
\end{figure}

Related to the concept of cyclomatic number is the {\em number of
spanning trees} of the graph. A spanning tree is a subset of the graph
that visits all nodes but has no loops (ie is a tree). A graph made up
from several disconnected parts has no spanning tree. A tree has
exactly one spanning tree. The number of spanning trees increases
rapidly with the cyclomatic number.

The height of the flattest spanning tree, or equivalently the maximum
number of hops separating two nodes on the graph (popularised in the
movie {\em six degrees of separation} --- which refers to the maximum
number of acquaintances connecting any two people in the World) is
another useful measure related to complexity. Networks having small
degrees of separation (so called {\em small world networks}) tend to support
more complex dynamics than networks having a large degree of
separation. The reason is that any local disturbance is propagated a
long way through a small world network before dying out, giving rise
to chaotic dynamics, whereas in the other networks, disturbances
remain local, leading to simpler linear dynamics.

\subsection{Offdiagonal complexity}

Recently, Claussen \citeyear(2007){Claussen04} introduced a complexity measure
called {\em offdiagonal complexity}
that is low for regular and randomly connected graphs, but takes on
extremal values for {\em scale-free} graphs, such as typically seen in
naturally occurring networks like metabolic and foodweb networks, the
internet, the world wide web and citation networks. The apparent
ubiquity of the scale-free property amongst networks we intuitively
associate as complex \cite{Newman03} is the justification for using
offdiagonal complexity, the other advantage being its computational
practicality.  

To compute offdiagonal complexity, start with the {\em adjacency
  matrix} 
\begin{equation}
  g_{ij} = \left\{
    \begin{array}{ll}
      1 &\mathrm{if}~i~\mathrm{and}~j~\mathrm{are~ connected}\\
      0&\mathrm{otherwise}\\
    \end{array}
  \right.
\end{equation}

Let $\ell(i)$ be the node degree of $i$, and let $c_{mn}, m\leq n$, be the number of edges between all nodes $i$ and $j$ with
node degrees $m=\ell(i)$, $n=\ell(j)$:
\begin{equation}
c_{mn}=\sum_{i,j} g_{ij}\delta_{m,\ell(i)}\delta_{n,\ell(j)} 
\end{equation}
Then the normalised $n$-th diagonal sum is
\begin{equation}
a_n=\frac{\sum_i c_{i,i+n}}{\sum_{i,j} c_{ij}}.
\end{equation}
The {\em offdiagonal complexity} is defined by a Boltzmann-Gibbs
entropy-like formula over the normalised diagonal sums:
\begin{equation}
C_{\mathrm{offdiag}} = -\sum_n a_n \ln a_n
\end{equation}

For regular lattices, each node has the same link degree, so $c_{mn}$
is diagonal, $a_n=\delta_{n0}$ and $C_{\mathrm{offdiag}}=0$. 

For random graphs, most edges will connect nodes with similar link
degree (the characteristic link degree scale), so $c_{mn}$ will have a
mostly banded structure, and $a_n\rightarrow0$ as $n$ increases. This
leads to small non-zero values of the offdiagonal complexity.

Scale free networks have a power law distribution of link degree,
which leads to the $c_{mn}$ matrix having a wide spread of entries. In
the case of all $a_n$ being equal, offdiagonal complexity takes its
maximum value as equal to the number of nodes.

\section{Information as Complexity}\label{info-complexity}

The single simplest unifying concept that covers all of the preceding
considerations is {\em information}. The more information required to
specify a system, the more complex it is. A sandpile is simple,
because the only information required is that it is made of sand
grains (each considered to be identical, even if they aren't in
reality), and the total number of grains in the pile. However, a
typical motorcar requires a whole book of blueprints in its
specification.

Information theory began in the work of Shannon
\citeyear(1949){Shannon49}, who was concerned with the practical problem of
ensuring reliable transmission of messages. Every possible message has
a certain probability of occurring. The less likely a message is, the
more information it imparts to the listener of that message. The
precise relationship is given by a logarithm:
\begin{equation}\label{shannon}
I = -\log_2 p
\end{equation}
where $p$ is the probability of the message, and $I$ is the
information it contains for the listener. The base of the logarithm
determines what units information is measured in --- base 2 means the
information is expressed in {\em bits}. Base 256 could be used to
express the result in {\em bytes}, and is of course equivalent to
dividing equation (\ref{shannon}) by 8.

\begin{figure}
\fbox{
\begin{pspicture}(0,1)(15,9)
\rput[l](0,5){\epsfxsize=8cm\epsfbox{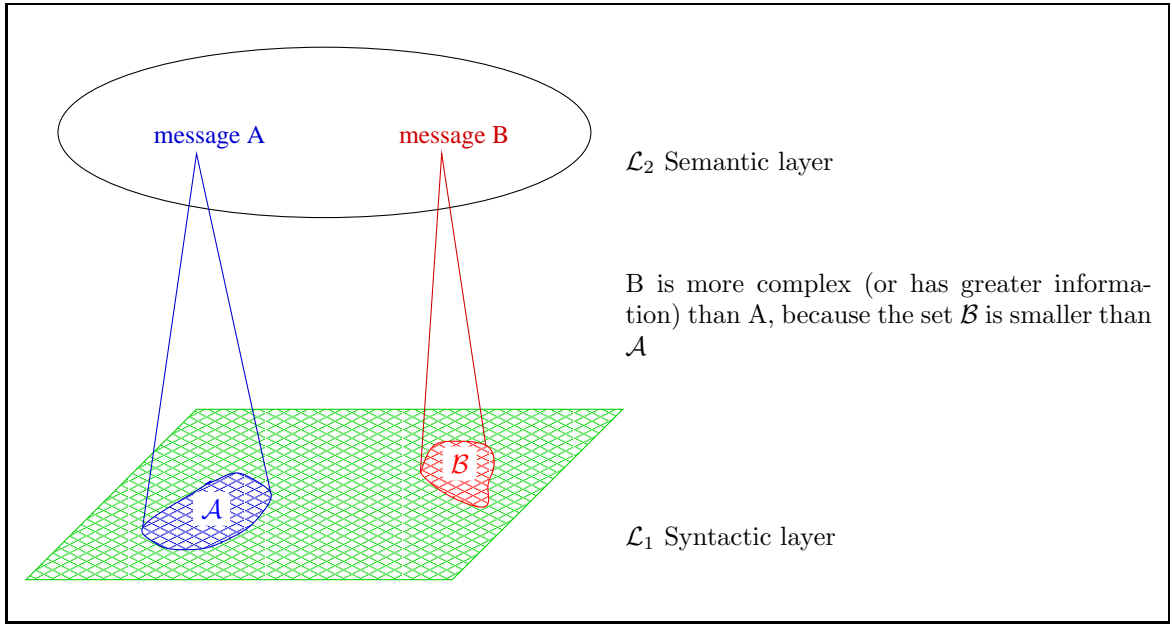}}
\rput(2.5,2.4){{\blue\psframebox*[framearc=.3]{${\cal A}$}}}
\rput(5.8,3){{\red\psframebox*[framearc=.3]{${\cal B}$}}}
\rput[l](8,2){${\cal L}_1$ Syntactic layer}
\rput[l](8,7){${\cal L}_2$ Semantic layer}
\rput[l](8,5){\parbox{7cm}{B is more complex (or has greater
    information) than A, because the set ${\cal B}$ is smaller
  than ${\cal A}$}}
\end{pspicture}
}
\caption{Diagram showing the syntactic and semantic spaces. Two
  different messages, having meanings A and B, can each be coded in
  many equivalent ways in syntactic space, represented by the sets
  ${\cal A}$ and ${\cal B}$. The information or complexity of the
  messages is related to the size it occupies in syntactic space by
  formula (\ref{shannon})}
\label{syn-sem}
\end{figure}

Shannon, of course, was not so interested in the semantic content of
the message (ie its meaning), rather in the task of information
transmission so instead considered a message composed of symbols $x_i$
drawn from an alphabet $A$. Each symbol had a certain probability
$p(x_i)$ of appearing in a message --- consider how the letter `e' is
far more probable in English text than the letter `q'. These
probabilities can be easily measured by examining extant texts. A first
order approximation to equation (\ref{shannon}) is given by:
\begin{equation} 
I(x_1x_2\ldots x_n) \approx \sum_{i=1}^n p(x_i)\log_2 p(x_i)
\end{equation}
This equation can be refined by considering possible pairs of letters,
then possible triplets, in the limit converging on the minimum amount
of information required to be transmitted in order for the message to
be reconstructed in its original form. That this value may be
considerably less that just sending the original message in its
entirety is the basis of compression algorithms, such as those
employed by the well-known {\em gzip} or {\em PKzip} (aka WinZip)
programs.

The issue of semantic content discouraged a lot of people from applying
this formalism to complexity measures. The problem is that a message
written in English will mean something to a native English speaker,
but be total gibberish to someone brought up in the Amazon jungle with
no contact with the English speaking world. The information content of
the message depends on exactly who the listener is! Whilst this
context dependence appears to make the whole enterprise hopeless, it
is in fact a feature of all of the measures discussed so far. When
counting the number of parts in a system, one must make a decision as
to what exactly constitutes a part, which is invariably somewhat
subjective, and needs to be decided by consensus or convention by the
parties involved in the discussion. Think of the problems in trying
decide whether a group of animals is one species of two, or which
genus they belong to. The same issue arises with the characterisation
of the system by a network. When is a relationship considered a graph
edge, when often every component is connected to every other
part in varying degrees.

However, in many situations, there appears to be an obvious way of
partitioning the system, or categorising it. In such a case, where two
observers agree on the same way of interpreting a system, then they
can agree on the complexity that system has. If there is no agreement
on how to perform this categorisation, then complexity is meaningless

To formalise complexity then, assume as given a classifier system
that can categorise descriptions into equivalence classes. Clearly,
humans are very good at this --- they're able to recognise patterns
even in almost completely random data. Rorschach plots are random ink
plots that are interpreted by viewers as a variety of meaningful
images. However, a human classifier system is not the only
possibility. Another is the classification of programs executed by a
computer by what output they produce. Technically, in these
discussions, researchers use a {\em Universal Turing Machine} (UTM),
an abstract model of a computer.

Consider then the set of possible binary strings, which can fed into a
UTM $U$ as a program. Some of these programs cause $U$ to produce some
output then halt. Others will continue executing forever. In
principle, it is impossible to determine generally if a program will
halt or continue on indefinitely. This is the so called {\em halting
problem}. Now consider a program $p$ that causes the UTM to
output a specific string $s$ and then halt. Since the UTM halts after
a certain number of instructions executed (denoted $\ell(p)$) 
the same result is produced by feeding in any string starting with the same
$\ell(p)$ bits.  If the strings have equal chance of being chosen
({\em uniform measure}), then the proportion of strings starting with the
same initial $\ell(p)$ bits is $2^{-\ell(p)}$. This leads to the {\em
universal prior} distribution over descriptions $s$, also known as the
Solomonoff-Levin distribution:
\begin{equation}\label{universal prior}
P(s) = \sum_{\{p: U(p)=s\}}2^{-\ell(p)}
\end{equation}

The complexity (or information content) of the description is given by
equation (\ref{shannon}), or simply the logarithm of (\ref{universal
prior}). In the case of an arbitrary classifier system, the
complexity is given by the negative logarithm of the equivalence class size
\begin{equation}\label{complexity}
{\cal C}(x) = \lim_{s\rightarrow\infty} s\log_2 N - \log_2 \omega(s,x)
\end{equation}
where $N$ is the size of the alphabet used to encode the description
and $\omega(s,x)$ is the number of equivalent descriptions having
meaning $x$ of size $s$ or less \cite{Standish01a}.

It turns out that the probability $P(s)$ in equation (\ref{universal
  prior}) is dominated by the shortest program \cite[Thm
4.3.3]{Li-Vitanyi97}, namely
\begin{equation}
K(s)+\log_2 P(s) \leq C
\end{equation}
($\log_2 P(s) <0$ naturally) where $C$ is a constant independent of
the description $s$. $K(s)$ is the length of the shortest program $p$
that causes $U$ to output $s$, and is called the {\em Kolmogorov
  complexity} or {\em algorithmic complexity}.

An interesting difference between algorithmic complexity, and the
general complexity based on human observers can be seen by considering
the case of random strings. {\em Random}, as used in algorithmic
information theory, means that no shorter algorithm can be found to
produce a string than simply saying ``print \ldots'', where the
\ldots{} is a literal representation of the string. The algorithmic
complexity of a random string is high, at least as high as the length
of the string itself. However, a human observer simply sees a random
string as a jumble of letters, much the same as any other random
string. In this latter case, the equivalence class of random strings
is very large, close to $N^s$, so the perceived complexity
is small. Thus the human classifier defines an example of what
Gell-Mann calls {\em effective complexity} \cite{Gell-Mann94}, namely a
complexity that has a high value for descriptions that are partially
compressible by complex schema, but low for random or obviously
regular systems.

A good introduction to information theoretical concepts for complex
systems studies can be found in \cite{Adami98a}.

\subsection{Information Theoretic Graph Complexity}

In an attempt to bridge the information theoretic approach to
complexity with graph theoretical approaches, Standish recently
introduced a coding scheme for which practical (though still NP-hard)
algorithms exist for calculating the size of the equivalence class of
descriptions \cite{Standish05a}. The intention is to use this method
with networks that have a meaning or function attached, such as
metabolic networks, or food webs. A randomly constructed food web will
collapse fairly quickly under ecosystem dynamics to a much smaller
stable food web, so random or regular networks will tend to have a lower
complexity value.

However, when applied to abstract networks, it leads to a perverse
result that regular networks have the high complexities, and the
completely connected network has maximal complexity. This effect can be
ameliorated by introducing a compressed complexity measure, which reduces
the complexity measure of regular networks, and is closer to a Turing
complete {\em syntactic} language. Unfortunately, there is no
computationally effective algorithm known for calculating this
compressed complexity measure.

\section{Computational Complexity and Logical Depth}

Algorithmic complexity takes no account of the time required to
execute the shortest algorithm. An almanac of tide charts and Newton's
equations of motion plus law of gravity for the Earth-Moon-Sun system contain
the same information, yet the almanac contains the information in a
more useful form for the maritime sailor, as it requires less work to
determine when the tides occur. {\em Computational complexity} and
{\em logical depth} are two concepts designed to
address the issue of a description's value.

{\em Computational Complexity} is defined as the execution time of the
algorithm. Since this is highly dependent on what operations are
available to the processor, usually only the scaling class of the
algorithm is considered as the input size is increased. Algorithms may
scale {\em polynomially}, which means the execution time increases as
some power of the problem size ($t\propto n^s$), or may scale faster
than this (eg exponentially: $t \propto s^n$), in which case they have {\em
nonpolynomial complexity}. The class of polynomial algorithms is
called $P$, and the class of nonpolynomial algorithms for which the
solution can be checked in polynomial time is called $NP$. It is known
that all $NP$ algorithms can be transformed into any other by means of
a polynomially preprocessor, but it is unknown whether or not $P=NP$,
ie whether it is possible to transform any nonpolynomial algorithm
into a polynomial one. This issue is of great importance, as certain
public key encryption schemes depend on nonpolynomial algorithms to
ensure the encryption scheme cannot be cracked within a practical
amount of time.

Bennett's {\em logical depth} \cite{Bennett88a} is the execution time
of the most highly compressed representation of an object, relative to
some reference machine. It is meant to be a measure of the value of
the object --- the almanac of tide tables has a high value of logical
depth compared with the equations of motion that generate it.

In terms of the observer-based complexity notions introduced in
\S\ref{info-complexity}, assume that the observer has a limited
amount of computing resources. Perhaps e is only prepared to spend
5 minutes computing the information, and prior to the widespread
availability of electronic computers this meant that the almanac was
not equivalent to the equations of motion, since it requires more than
5 minutes to compute the tidal information from the equations of
motion via manual paper and pencil techniques. Since the almanac is
inequivalent to the equations of motion {\em in this context}, it is
clear that the almanac has greater complexity {\em in this context}.

\section{Occam's Razor}

The practice of preferring a simpler theory over a more complex one
when both fit the observed evidence is called {\em Occam's Razor},
after William de Occam:
\begin{quote}
Entities should not be multiplied unnecessarily.
\end{quote}

What is not widely appreciated, is that this strategy is remarkably
successful at picking better theories. Often, when tested against
further empirical evidence, the simpler theory prevails over the more
complex. A classical example of this sort of thing is Einstein's
General Theory of Relativity. The key field equations of General
Relativity are really quite simple:
\begin{equation}
{\sf\bf G}=8\pi\kappa{\sf\bf T}
\end{equation}
Of course unraveling what these equations means for a specific
instance involves a nontrivial amount of 4 dimensional tensor
calculus, so the General Relativity computations have high logical
depth. Einstein proposed the equations in this form because they
seemed the most ``beautiful'' --- there were a large number of
alternative formulations that fitted the data at the time. One by one,
these alternative formulations were tested empirically as technology
developed through the 20th century, and found wanting.

However, by what criteria is a particular theory more simple than
another. Goodman \citeyear(1972){Goodman72} developed a theory of {\em
  simplicity} to put the practice of Occam's Razor on a more rigorous
footing. His idea was to formalise the theories into formal logic
predicates, and then count the number of primitive clauses required to
encode the theory.

Solomonoff \citeyear(1964){Solomonoff64} developed the concept of {\em
  algorithmic information complexity} in the 1960s as a way of
explaining why Occam's razor works. He considered the set of all
possible descriptions and computed the probability distribution that a
particular description would be generated by a program picked at
random by the reference machine. His work had some technical problems
that were solved by Levin \citeyear(1974){Levin74}, which led
to the {\em universal prior} distribution (\ref{universal prior}).
Basically, simple descriptions have a much higher probability than
more complex ones, thus Occam's razor. The same behaviour is true of
the arbitrary classifier system at equation
(\ref{complexity}) \cite{Standish00a}.

While the world is expected to be remarkably simple by the above
arguments, it is also logically very deep ($10^{10}$ years of runtime
so far!). This appears to be the result of another poorly understood
principle called the {\em Anthropic Principle}. The Anthropic
Principle states that the world must be consistent with our existence
as intelligent, reasoning beings \cite{Barrow-Tipler86}. So while
Occam's razor says we should live in the simplest of universes, the
Anthropic Principle says it shouldn't be too simple, as a certain
level of complexity is required for intelligent life. The simplest
means of generating this level of complexity is by accruing random
mutations, and selecting for functional competence, ie Darwinian evolution.

\section{Complexity as a quality --- Emergence}

It is often thought that {\em complex systems} are a separate category
of systems to {\em simple systems}. So what is it that distinguishes a
complex system, such as a living organism, or an economy, from a
simple system, such as a pair of pliers? This question is related to
the notorious question of {\em What is Life?}, however may have a
simpler answer, since not all complex systems are living, or even
associated with living systems.

Consider the concept of {\em emergence} \cite{Holland97,Fromm04}. We
intuitively recognise emergence as patterns arising out of the
interactions of the components in a system, but not implicit in the
components themselves. Examples include the formation of hurricanes
from pressure gradients in the atmosphere, crashes in stock markets,
flocking behaviour of many types of animals and of course, life
itself.

Let us consider a couple of simple illustrative examples, that are
well known and understood. The first is the {\em ideal gas}, a model
gas made up of large numbers of non-interacting point particles
obeying Newton's laws of motion. A {\em thermodynamic} description of
the gas is obtained by averaging: 
\begin{description}
\item[temperature ($T$)] is the average kinetic energy of the particles;
\item[pressure ($P$)] is the average force applied to a unit area of the
boundary by the particles colliding with it;
\item[density ($\rho$)] is the average mass of particles in a unit volume;
\end{description}
The {\em ideal gas law} is simply a reflection of the underlying laws
of motion, averaged over all the particles:
\begin{equation}
P\rho\propto T
\end{equation}
The thermodynamic state is characterised by the two parameters $T$ and
$\rho$. The so-called {\em first law of thermodynamics} is simply a
statement of conservation of energy and matter, in average form.

An entirely different quantity enters the picture in the form of {\em
entropy}. Consider discretising the underlying phase-space into cubes
of size $h^N$, ($N$ being the number of particles) and then counting
the number of such cubes having temperature $T$ and density $\rho$,
$\omega(T,\rho,N)$. The entropy of the system is given by
\begin{equation}
S(T,\rho,N)=k_B \ln \omega(T,\rho,N)
\end{equation}
where $k_B$ is a conversion constant that expresses entropy in units
of Joules per Kelvin.  One can immediately see the connection between
complexity (eq. \ref{complexity}) and entropy. Readers familiar with
quantum mechanics will recognise $h$ as being an analogue of Planck's
constant. However, the ideal gas is not a quantum system, and as
$h\rightarrow0$, entropy diverges! However, it turns out that in the
thermodynamic limit ($N\rightarrow\infty$), the average entropy $S/N$
is independent of the size of $h$.

The {\em second law of thermodynamics} is a recognition of the fact
that the system is more likely to move a state occupying a larger
region of phase space, than a smaller region of phase space, namely
that $\omega(T,\rho,N)$ must increase in time. Correspondingly entropy
must also increase (or remain constant) over time. This is a
probabilistic statement that only becomes exact in the thermodynamic
limit. At the syntactic, or specification level of description (ie
Newton's laws of motion), the system is perfectly {\em reversible} (we
can recover the system's initial state by merely reversing the
velocities of all the particles), yet at the semantic (thermodynamic)
level, the system is {\em irreversible} (entropy can only increase, never
decrease).

The property of irreversibility is an {\em emergent} property of the
ideal gas, as it is not {\em entailed} by the underlying
specification. It comes about because of the additional identification
of thermodynamic states, namely the set of all micro-states possessing
the same temperature and density. This is extra information, which in
turn entails the second law.

The second example I'd like to raise (but not analyse in such great
depth) is the well known {\em Game of Life}, introduced by John
Conway \citeyear(1982){Conway82}. This is a {\em cellular
  automaton} \cite{Wolfram84}, in this case, a 2D grid of cells where
each cell can be one of two states.  Dynamics on the system is imposed
by the rule that the state of a cell depends on the values of its
immediate neighbours at the previous time step.

Upon running the Game of Life, one immediately recognises a huge
bestiary of emergent objects, such as blocks, blinkers and gliders.
Take gliders for example. This is a pattern that moves diagonally
through the grid. The human observer recognises this pattern, and can
use it to {\em predict} the behaviour of the system with less effort
than simulating the full cellular automaton. It is a {\em model} of
the system. However, the concept of a glider is not {\em entailed} by
the cellular automaton specification, which contains only states and
transition rules. It requires the additional identification of a
pattern by the observer.

This leads to a general formulation of {\em
  emergence} \cite{Standish01a}. Consider a system specified in a
language ${\cal L}_1$, which can be called the specification, or {\em
  syntactic} layer (see figure \ref{syn-sem}). If one accepts the
principle of {\em reduction}, all systems can ultimately be specified
the common language of the theoretical physics of elementary
particles. However, an often believed corollary of reduction is that
this specification encodes all there is to know about the system. The
above two examples shows this corollary to be manifestly false. Many
systems exhibit one or more {\em good models}, in another language
${\cal L}_2$, which can be called the {\em semantic layer}. The
system's specification does not entail completely the behaviour of the
semantic model, since the latter also depends on specific
identifications made by the observer. In such a case, we say that
properties of the semantic model is emergent with respect to the
syntactic specification.

The concept of ``good'' model deserves further discussion. In our
previous two examples, neither the thermodynamic model, nor the glider
model can be said to perfectly capture the behaviour of the
system. For example, the second law of thermodynamics only holds in
the thermodynamic limit --- entropy may occasionally decrease in finite
sized systems. A model based on gliders cannot predict what happens
when two gliders collide. However, in both of these cases, the
semantic model is cheap to evaluate, relative to simulating the full
system specification. This makes the model ``good'' or ``useful'' to
the observer. We don't prescribe here exactly how to generate good
models here, but simply note that in all cases of recognised emergence, the
observer has defined a least one semantic and one syntactic model of the
system, and that these models are fundamentally
incommensurate. Systems exhibiting emergence in this precise sense can
be called {\em complex}.

A school of thought founded by Robert Rosen holds that complex systems
cannot be described by a single best model as reductionists would
assume, but rather has a whole collection of models that in the limit
collectively describe the system \cite{Rosen91}. That such systems
exist, at least formally, is assured by G\"odel's incompleteness
theorem, \cite{Hofstadter79} which shows that number theory is just
such a system that cannot be captured by a finite specification. He
further argues mechanical systems (those that have a finite
specification such as the examples I have given above) can never be
complex, since the specification contains all there is to know about
the system. However, he implicitly assumes that all models must be
perfect (ie in perfect correspondence with the underlying system),
rather than merely good as I do here. This constitutes a {\em straw
  man} argument, and leads him to the {\em false} conclusion that
mechanical systems (eg computer simulations) can never exhibit
emergence. The two examples presented above, which are perfectly good
mechanical systems, are counter-examples to this claim. Furthermore,
the definition of complex systems presented here is known to be
non-empty, a fact not known of Rosen's definition since no physical
counterpart to G\"odel's incompleteness theorem is known.

\section{Conclusion}

When connoting a quality, complexity of a system refers to the
presence of emergence in the system, or the exhibition of behaviour
not specified in the systems specification. When connoting a quality,
complexity refers to the amount of information needed to specify the
system. Both notions are inherently observer or context dependent,
which has lead to a disparate collection of formalisations for the
term, and has lead to some despairing of the concept being adequately
formalised. This would be a mistake, as within a given application,
the meaning can be well defined.

An additional difficulty is the combinatorial size of the underlying
syntactic space, which can lead to the intractability of computing
complexity. Furthermore, the details of the syntactic layer may be
inaccessible, for example the absence of historical genetic data in
the study of evolution from the fossil record. So being able to
establish easy to measure proxies for complexity is often important,
and many proposals for complexity are of this nature.

\section{Bibliography}

\bibliographystyle{apalike}
\bibliography{rus}

\section*{Glossary}

\begin{description}
\item[Algorithmic Complexity] Length of shortest program capable of
generating the description (also known as Kolmogorov complexity)
\item[Anthropic Principle] The statement that the properties of the
universe must be such as to permit our existence as human
observers
\item[Cellular Automaton] A grid of cells, each of which can be in a
finite number of states. The states of each cell depend only on the
states of a neighbourhood of cells at the previous timestep, known as
a transition rule.
\item[Church-Turing Thesis] The proposition that all computable
functions can be computed on a Turing machine.
\item[Classifier system] A system that classifies a set of objects into a
discrete set of classes. Formally equivalent to a map into the set of
whole numbers.
\item[Complexity (quality)] The quality of possessing emergent properties
\item[Complexity (quantity)] The amount of information a particular system
represents to the observer
\item[Computational Complexity] The computational cost of executing an
algorithm. 
\item[Emergence] The phenomenon of emergent properties
\item[Emergent Properties] Properties of a system at the semantic
level that are not entailed at the syntactic level
\item[Entail] To logically imply something  
\item[Entropy] Logarithm of number of syntactic states equivalent to a
given semantic state. It is related to information ($I$) by
$S+I=S_\mathrm{max}$ where $S_\mathrm{max}$ is the maximum possible
entropy of the system.
\item[Equivalence Class] A set of objects that are equivalent under
some mapping i.e. $\{x : e(x)=e(x')\} \exists x'$
\item[Game of Life] A well known cellular automaton, with two states
per cell, and a particular transition rule
\item[Graph Theory] Mathematical theory of objects consisting of
atomic nodes linked by connections called edges
\item[G\"odel Incompleteness Theorem] No finite set of axioms can
prove all possible true theorems of number theory
\item[Information] The amount of meaning in any description; formally
given as the logarithm of the proportion of syntactic space occupied
by the description
\item[Link degree] the number of edges a node has in a graph  
\item[Logical Depth] Execution time of the most compressed
representation of an object
\item[Newton's Laws of motion] Laws of ideal point particles:
the acceleration a particle experiences is proportional to the force
acting on it, which is a function of the positions and velocities of
the particle and the environment
\item[Occam's Razor] A statement of the practice of preferring a
simpler theory over a more complex one
\item[Scale free] A scale free distribution has infinite mean. A power
  law distribution is a common scale free distribution. A scale
  free process is a stochastic process obeying a scale free distribution.
\item[Second Law of Thermodynamics] Entropy can only increase, or
remain constant in a closed system; it can never decrease
\item[Semantic Level (or space)] The space of meanings for any description.
\item[Syntactic Level (or space)] The language in which a description
is specified: letters of the alphabet, genetic code, laws of
theoretical physics, as appropriate
\item[Turing Machine] A formal model of a computation.
\item[Universal Turing Machine] A formal model of a
computer. Is capable is simulating any Turing Machine with appropriate input. 
\end{description}

\end{document}